\documentclass[aps,twocolumn,prd,superscriptaddress,nofootinbib,floatfix]{revtex4-2}

\usepackage{mathtools}
\usepackage{amsfonts}
\usepackage{amssymb}
\usepackage{mathrsfs}
\usepackage{bbm}
\usepackage{slashed}
\usepackage{amsmath}
\usepackage{bm}
\usepackage{bbold}

\usepackage{graphicx}
\usepackage{color}
\usepackage{colortbl}
\usepackage{array}

\usepackage{float}
\usepackage{placeins}
\usepackage{booktabs}
\usepackage[caption=false]{subfig}
\usepackage{makecell}
\usepackage{tabstackengine}

\usepackage{xspace}
\usepackage{hyperref}
\usepackage[nameinlink]{cleveref}
\usepackage{bookmark}
\usepackage{siunitx}

\usepackage{xifthen}
\usepackage{xcolor}
\hypersetup{
	colorlinks,
	linkcolor={red!75!black},
	citecolor={blue!75!black},
	urlcolor={blue!75!black}
}

\usepackage[utf8]{inputenc}
\allowdisplaybreaks[4]
\usepackage[normalem]{ulem}

\newcommand{\feyn}[1]{
	\setbox0=\hbox{\ensuremath{#1}}
	\hbox to\wd0{\hbox to0pt{\hbox to\wd0{\hss/\hss}\hss}\box0}}

\def\Eq#1{Eq.~\labelcref{#1}}

\def\eq#1{\labelcref{#1}}
\def\Fig#1{Fig.~\labelcref{#1}}

\def\sec#1{Sec.~\labelcref{#1}}

\setkeys{Gin}{width=0.48\textwidth}
\graphicspath{{./figures/}}

\newcommand{\gettitle}{}

\newcommand{\getLanzhouAffiliation}{\affiliation{School of Physical Science and Technology, Lanzhou University, Lanzhou 730000, China}}

\newcommand{\getDalianAffiliation}{\affiliation{School of Physics, Dalian University of Technology, Dalian, 116024, P.R. China}}

\newcommand{\getGiessenAffiliation}{\affiliation{Institut f\"ur Theoretische Physik, Justus-Liebig-Universit\"at Gie\ss en, 35392 Gie\ss en, Germany}}

\newcommand{\getEnergyAffiliation}{\affiliation{Energy Singularity Fusion Power Technology (SH) Ltd., Shanghai, China}}

\hypersetup{
	colorlinks,
	linkcolor={red!75!black},
	citecolor={blue!75!black},
	urlcolor={blue!75!black},
	pdftitle={\gettitle},
	pdfauthor={Zelle},
	pdfkeywords={effective theory} {analytic continuation}
	{correlations functions} {hadronization}
	{functional renormalisation group}
	{real time} {bound states}
	bookmarksopen=true,
	bookmarksopenlevel=2,
	bookmarksnumbered=true
}

\begin{document}

\title{Deuteron yields near the QCD phase transition}
	
\author{Sheng-nan Han}
\getLanzhouAffiliation
\getDalianAffiliation

\author{Jing Wu}
\getDalianAffiliation

\author{Yong-rui Chen}
\getGiessenAffiliation

\author{Yi-zhen Huang}
\getDalianAffiliation

\author{Feng Li}
\email{lifengphysics@gmail.com}
\getEnergyAffiliation

\author{Wei-jie Fu}
\email{wjfu@dlut.edu.cn}
\getDalianAffiliation

\begin{abstract} 

We investigate the influence of QCD phase transition and critical fluctuations of the critical end point (CEP) on the deuteron yield within the functional renormalization group (fRG) approach, by using the nucleon coalescence model and a low energy effective field theory of quarks and mesons. It is found that the two-point baryon density correlation function is enhanced in a narrow region radiated from the CEP along the phase boundary. The deuteron yield arising from the two-point baryon correlation is small compared to the leading-order contribution, which is attributed to the fact that in the regime of low collision energy, i.e., the region of large baryon chemical potential, the freeze-out curves deviate from the critical region, resulting in that the enhancement of the deuteron yield stemming from the critical fluctuations near the CEP is mild.

\end{abstract}
	
\maketitle

\section{Introduction}
\label{sec:intro}

Quark-gluon plasma (QGP), a novel state of matter, is experimentally produced in high energy heavy-ion collisions under extreme high temperature and/or densities \cite{Shuryak:1980tp, Harris:1996zx, STAR:2005gfr, PHENIX:2004vcz, ALICE:2005vhb, Busza:2018rrf, STAR:2023jdd}. Under these extreme conditions, colored degrees of freedom are deconfined from hadrons, and thus the system transitions from the hadronic phase to a QGP phase dominated by quarks and gluons, among which the interactions are described by quantum chromodynamics (QCD). Moreover, this deconfinement phase transition is accompanied by the chiral phase transition, where the dynamically broken chiral symmetry in the hadronic phase, which accounts for the mass production of hadrons \cite{Fu:2022uow, Fu:2024ysj, Fu:2025hcm}, is restored in the QGP phase. As a novel state of matter, QGP has attracted extensive research interest due to its exceptional properties under extreme conditions. For instance, QGP exhibits near-perfect fluid behavior characterized by exceptionally low viscosity \cite{Voloshin:2008dg, Heinz:2013th, Song:2017wtw, Ollitrault:2023wjk}, resulting in extensive studies of its transport coefficients, such as the shear viscosity, bulk viscosity, and diffusion constants \cite{Feng:2017tsh, Bernhard:2019bmu}. 

Both universal and non-universal properties of QCD phase transitions are encoded in the structure of QCD phase diagram in the plane of, e.g., the temperature and baryon chemical potential. By elucidating the QCD phase diagram, one is able to unravel more essential properties of the nonperturbative nature of QCD. On the QCD phase diagram there might be a critical end point (CEP) \cite{Stephanov:1998dy, Stephanov:1999zu}, which is the end point of the first-order phase transition line in the regime of large baryon chemical potential, or baryon densities. Due to the universal property of phase transitions, the end point of the first-order phase transition line must be a second-order phase transition point of $Z(2)$, i.e., Ising universality class. By contrast, the existence or the location of CEP is a non-universal property, that is closely related to the specific essence of QCD. Remarkably, recent theoretical studies of QCD phase diagram have seen significant progress, see, e.g., \cite{Stephanov:2007fk, Fu:2019hdw, Gao:2020fbl, Gunkel:2021oya, Fu:2022gou}.

The search of the CEP is under way in heavy-ion collision experiments \cite{STAR:2010vob, Luo:2017faz, Andronic:2017pug, Bzdak:2019pkr, Chen:2024aom}. Since the correlation length increases significantly near a second-order phase transition point, fluctuation observables are employed to trace remnants of critical fluctuations related to the critical end point in experiments, e.g., the ripples of CEP \cite{Fu:2023lcm}. The fluctuation observables include the fluctuations and correlations of conserved charges \cite{STAR:2020tga, STAR:2021fge, STAR:2022vlo, STAR:2022etb, STAR:2025zdq, Fu:2015naa, Fu:2015amv, Fu:2016tey, Fu:2021oaw, Fu:2023lcm, Lu:2025cls}, such as the net-baryon or net-proton number, the electric charge, the strangeness, etc. Recently, there have also been efforts to try to use jets as a probe to explore the CEP \cite{Wu:2022vbu}. 

In recent years, the yields of light nuclei produced in relativistic heavy-ion collisions have attracted significant attention due to their unique sensitivity to the hadronization and freeze-out dynamics in heavy-ion collision \cite{Chen:2018tnh, Sun:2020zxy}. In fact, the yields of light nuclei are closely related to $n$-point correlation functions of baryon number. For instance, the yield of deuteron receives contribution from two-point baryon correlation function. Consequently, it is interesting to investigate the influence of critical fluctuations or their remnants related to the CEP on the yields of light nuclei.

In this work we would like to combine the nucleon coalescence model \cite{Sato:1981ez, Mrowczynski:1987oid, Scheibl:1998tk, Chen:2003ava, Blum:2019suo} and a low energy effective field theory (LEFT) within the functional renormalization group (fRG) approach to study the influence of QCD phase transition and critical fluctuations of CEP on the deuteron yields. The two-point baryon correlation function will be obtained from its flow equation, that is consistent with recent results of QCD phase diagram \cite{Fu:2019hdw, Gao:2020fbl, Gunkel:2021oya, Fu:2023lcm}. The fRG provides us with a systematic method to compute the correlation functions of different orders in a nonperturbative way, and thus is tailor-made for studies of QCD phase transitions, see e.g., \cite{Fu:2019hdw, Braun:2020ada, Chen:2021iuo, Braun:2023qak, Tan:2024fuq, Fu:2024rto, Chen:2025vwl, Dupuis:2020fhh, Fu:2022gou} for recent progress.

This paper is organized as follows: In \sec{sec:yield} we illustrate the relation between the deuteron yield and the two-point baryon density correlation function in the nucleon coalescence model firstly, and then calculate the baryon correlation function in a LEFT within the fRG approach. Numerical results are presented in \sec{sec:num}. We give a summary with conclusions in \sec{sec:con}. The explicit expressions of flows of the two-point baryon correlation function are collected in \Cref{app:flow-2point}.

\section{The Deuteron Yield}
\label{sec:yield} 

\subsection{Theoretical basis}

The nucleon coalescence model, a phenomenological framework widely employed to describe light nucleus production in heavy-ion collisions, incorporates both nucleon-nucleon correlations and phase-space distributions \cite{Sato:1981ez, Mrowczynski:1987oid, Scheibl:1998tk, Chen:2003ava, Blum:2019suo}. Based on this framework, as shown in Ref.~\cite{Sun:2020zxy}, the deuteron yield can be expressed as
\begin{align}
    N_d\approx{N_d^{(0)}(1+C_{np})+N_d^{(C_2)}}\,,\label{eq:Nd}
\end{align}
with
\begin{align}
    N_d^{(0)}=6\Big(\frac{2\pi}{mT}\Big)^{3/2}\frac{\sigma_d^3}{\big(\frac{1}{mT}+2\sigma_d^2\big)^{3/2}}N_p\langle\rho_n\rangle\,,\label{eq:Nd0}
\end{align}
standing for the deuteron yield without density fluctuations or correlations in the emission source based on the usual coalescence model, where the parameter $\sigma_d$ indicates the width of deuteron, which is related to the root-mean-square radius $r_d$ of deuteron via $\sigma_d=\sqrt{3/4}~r_d\approx2.26~\mathrm{fm}$ \cite{Ropke:2008qk}. \Cref{eq:Nd} is derived at the freeze-out stage and should therefore be evaluated with the freeze-out temperature $T$, the fireball volume $V$ and the nucleon mass $m$. Here,
\begin{align}
    C_{np}=\frac{\langle\delta\rho_n(x)\delta\rho_p(x)\rangle}{\langle\rho_n\rangle\langle\rho_p\rangle}\,,\label{eq:Cnp}
\end{align}
represents the correlation between the neutron and proton density fluctuations. Here $N_p$ denotes the proton yield in the emission source, $\rho_n$ and $\rho_p$ the neutron and proton densities, respectively. The symbol $\langle\cdots\rangle$ denotes the average over the coordinate space. $C_{np}$ in \Eq{eq:Cnp} can be regarded as a small parameter in the current framework. 

$N_d^{(C_2)}$ in \Eq{eq:Nd} represents the modification of deuteron yield arising from the neutron and proton density correlation, which reads
\begin{align}
    N_d^{(C_2)}\approx \frac{3}{\sqrt{2}}\Big(\frac{1}{2}+mT\sigma_d^2\Big)^{-\frac{3}{2}}V\bar C_2\,,\label{eq:Ndc2}
\end{align}
with 
\begin{align}
    \bar C_2=\int{d^3{\bm{x}}C_2(\bm{x})}e^{-\frac{\bm{x}^2}{2\sigma_d^2}}\,,\label{eq:C2bar}
\end{align}
where $C_2(\bm{x})$ represents the equal-time correlation function between the neutron and proton density fluctuations. 
In contrast to the leading-order contribution in \Eq{eq:Nd0}, the nonlocal second-order fluctuation of the baryon density $\bar C_2$, related to the neutron and proton density fluctuation correlation function $C_2(\bm{x})$ as shown in \eq{eq:C2bar}, can be extended to characterize correlations that originate at earlier time. This arises from the fact that the single-particle distribution is expected to be close to local thermal equilibrium at kinetic freeze-out and thus carries little memory about earlier evolution, whereas $C_2(\bm{x})$ typically exhibits a longer relaxation time. Consequently, $\bar C_2$ more probably keep remnants of fluctuations that emerged near the chemical freeze-out, or even during the phase transition.

In our theoretical calculations, we adopt a simplified treatment by neglecting the isospin dependence of $C_2(\bm{x})$ and directly employing it as the equal-time two-point baryon density correlation function. Since the real-time and Matsubara equal-time Green's functions are identical, we will perform the following calculations in the Matsubara formalism. We define the Matsubara two-baryon density fluctuation correlation function and its Fourier transform as:
\begin{align}
    C_2(\tau_1-\tau_2,\bm{x}_1-\bm{x}_2)\equiv\langle \delta j^0(-\mathrm{i}\tau_1,\bm{x}_1),\delta j^0(-\mathrm{i}\tau_2,\bm{x}_2)\rangle\,,\label{eq:c2}
\end{align}
and
\begin{align}
    C_2(p_0,\bm{p})\equiv \int_{0}^{1/T} d{\tau}\int d^3\bm{x}\, C_2(\tau,\bm{x}) e^{-\mathrm{i}(p_0 \tau+\bm{p}\cdot\bm{x})}\,,\label{}
\end{align}
with the baryon number density operator $j^0(x)=\frac{1}{3}\bar{q}(x)\gamma^0 q(x)$ and $\delta j^0=j^0-\langle j^0\rangle$, where $p_0=2\pi nT~(n\in Z)$ is the Matsubara frequency and $q(x)$ stands for the quark field. Then the equal-time two-baryon density fluctuation correlation can be given by
\begin{align}
    C_2(\tau=0,\bm{x})=T\sum_n \int \frac{d^3\bm{p}}{(2\pi)^3}\, C_2(p_0,\bm{p}) e^{\mathrm{i}\bm{p}\cdot\bm{x}}\,.\label{eq:C2x}
\end{align}
Substituting \Eq{eq:C2x} into \Eq{eq:C2bar}, one arrives at
\begin{align}
    \bar{C}_2 =\sqrt\frac{2}{\pi}\sigma_d^3 \int_0^{\infty} d{|\bm{p}|} \, {|\bm{p}|}^2 C_2(\tau=0,|\bm{p}|) e^{-\frac{1}{2} {|\bm{p}|}^2 \sigma_d^2}\,.\label{eq:C2barII}
\end{align}
One can see that $C_2(\tau=0,|\bm{p}|=\sqrt{2}/\sigma_d)$ contributes most to the integral above. In the following we will compute the two-point baryon density correlation in a low-energy effective model within the fRG approach.

\subsection{Two-point baryon density correlations}

In this work we adopt the quark-meson (QM) low-energy effective theory used in our previous work \cite{Fu:2021oaw, Fu:2023lcm}. The QM model provides a useful tool for exploring QCD properties at finite temperature and chemical potential. This model describes the interactions between quarks and mesons which govern the dynamics in the low-energy region. Since the QCD phase transition is predominantly driven by the quarks and mesons of light flavors, here we employ a two-flavor QM model, comprising the two light flavor quarks, the scalar sigma and the pseudo-scalar pion mesons. It has been already shown in previous studies that the QM model can well describe the chiral phase transition, see, e.g., \cite{pawlowski:2014zaa, Schaefer:2004en}. The strange quark is heavier than the light quarks, and thus we do not take it into account in the current computation, since it produces a subleading contribution. Moreover, the gluon dynamics plays a role in the regime of high energy, which is not included in the low-energy effective theory. 

We begin with the Euclidean effective action of the QM model as follows
\begin{align}
    \Gamma_k[\Phi]&=\int_{0}^{1/T} d{\tau}\int d^3\bm{x} \Big\{\bar{q}\Big[\gamma_\mu \partial_\mu-\big(\mu+\mu'(x)\big) \gamma_0\Big]q\nonumber\\[2ex]
    &\hspace{-0.3cm}+\frac{1}{2}\big(\partial_\mu \phi \big)^2+h_k\bar{q}\big(T^0\sigma+i \gamma_5\bm{T} \cdot \bm{\pi}\big)q +V_k(\rho)-c\sigma\Big\}\,,\label{eq:action}
\end{align}
with the fields $\Phi=(q, \bar q, \phi)$, i.e., the $u$, $d$ two-flavor quark field $q=(q_u, q_d)^{\top}$ and the sigma and pion meson fields $\phi=(\sigma, \bm\pi)$. The subscript $k$ represents the infrared (IR) cutoff, i.e., the renormalization group (RG) scale. The generators of $U(2)$ group in the flavor space read $(T^0,\bm T)=1/2 (\mathbb{1}, \bm \sigma)$ with the Pauli matrices $\bm \sigma$. The quarks interacts with the mesonic fields through the Yukawa coupling $h_k$. The mesonic effective potential $V_k(\rho)$ with $\rho=\phi^2/2$ in \Eq{eq:action} is $O(4)$ invariant. The last term in \Eq{eq:action} with the strength $c$ breaks the chiral symmetry explicitly.

In \Eq{eq:action} $\mu$ is the quark chemical potential, related to the baryon chemical potential through $\mu=\mu_B/3$. Note that the isospin symmetry is assumed for the $u$ and $d$ quarks and they have the same chemical potential $\mu$. The coordinate dependent $\mu'(x)$ can be regarded as the external source coupled with the density operator $\bar{q}\gamma_0 q$, which is introduced in \Eq{eq:action} for the convenience of computation of the density correlation in \Eq{eq:c2}.

The expectation value of the $\sigma$ field serves as the order parameter of chiral phase transition. At high baryon chemical potential, its value exhibits a discontinuous jump, corresponding to a first-order phase transition, whereas at low baryon chemical potential it varies smoothly in a crossover. The continuous crossover is connected with the first-order phase transition line at the critical end point (CEP), which is well described by the QM effective action in \Eq{eq:action}.

%
\begin{figure}[t]
\includegraphics[width=0.48\textwidth]{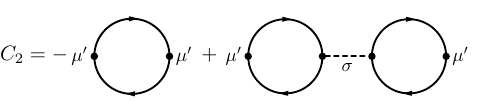}
\caption{Diagrammatic representation of the two-baryon density correlation. Black dots denote vertices; the solid and dashed lines represent the quark and sigma propagators, respectively.}
\label{fig:ogogsg}
\end{figure}
%

%
\begin{figure}[t]
\includegraphics[width=0.26\textwidth]{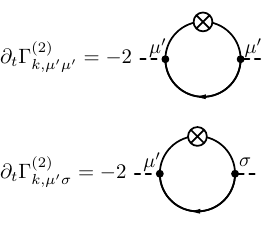}
\caption{Flow equations for the two-point functions $\Gamma^{(2)}_{k,\mu'\mu'}$ and $\Gamma^{(2)}_{k,\mu'\sigma}$ in \eq{eq:d2Gamd2mu,eq:d2Gamdmudsigma}. The black dots denote vertices, and the crossed circle represents the IR regulator in the fRG. The solid lines represent the quark fields. The dashed lines denote the external field $\mu'$ or the $\sigma$ field.}\label{fig:op2}
\end{figure}
%

%
\begin{figure*}[t]
\includegraphics[width=0.45\textwidth]{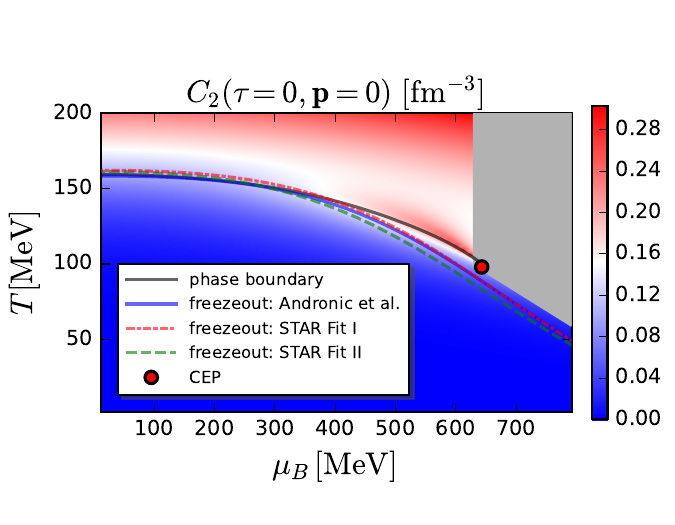}\hspace{0.3cm}
\includegraphics[width=0.45\textwidth]{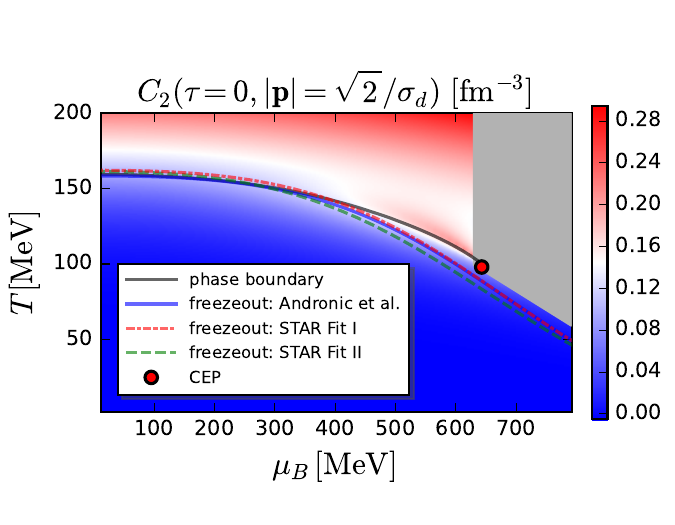}
\caption{Heatmap of the two-point baryon density correlation $C_2(\tau=0,\bm{p}=0)$ (left panel) and $C_2(\tau=0,|\bm{p}|=\sqrt{2}/\sigma_d)$ (right panel) in the QCD phase diagram. The red dot denotes the critical end point. The black solid line represents the phase boundary of chiral crossover, and other lines represent three different freeze-out curves \cite{Fu:2021oaw}. The gray area depicts the region where the computation is inaccessible yet.}\label{fig:C2}
\end{figure*}
%

%
\begin{figure}[t]
\includegraphics[width=0.45\textwidth]{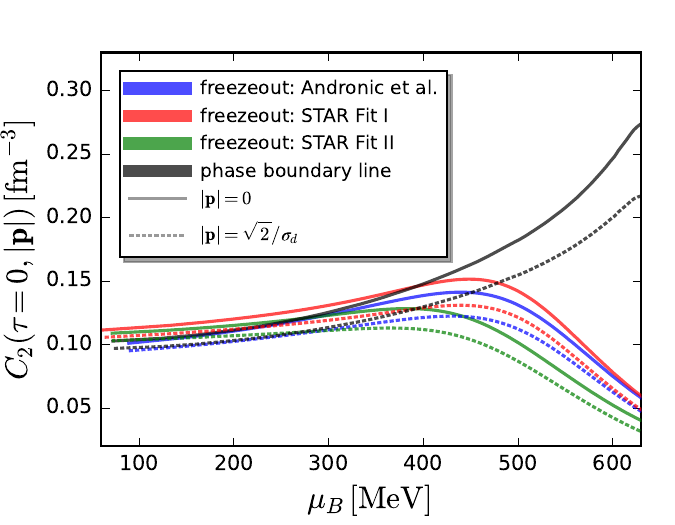}
\caption{Two-point baryon density correlation as a function of the baryon chemical potential on the phase boundary line and the three freeze-out curves. The solid and dashed lines correspond to $\bm{p}=0$ and $|\bm{p}|=\sqrt{2}/\sigma_d$, respectively.}
\label{fig:C2-muB}
\end{figure}
%

Then, the two-point baryon density correlation in \Eq{eq:c2} can be computed from the effective action in \Eq{eq:action} by taking twice derivatives with respect to $\mu'$, i.e.,
\begin{align}
    C_2(x_1-x_2)=&- \frac{1}{9}\frac{\delta^2\Gamma_{k}[\Phi (\mu'), \mu']}{ \delta \mu'(x_1) \delta\mu'(x_2)}\Big|_{\mu'=0}\,,\label{eq:C2-d2Gamd2mu}
\end{align}
which can be decomposed into a sum of two parts after a brief calculation, that is,
\begin{align}
    C_2(x_1-x_2)=&-\frac{1}{9}\Big(\Gamma^{(2)}_{k,\mu'\mu'}-\Gamma^{(2)}_{k,\mu'\sigma}G_{\sigma}\Gamma^{(2)}_{k,\mu'\sigma}\Big)\,,\label{eq:C2-twoparts}
\end{align}
with 
\begin{align}
    \Gamma^{(2)}_{k,\mu'\mu'}=&\frac{\delta^2\Gamma_{k}[\Phi, \mu']}{ \delta \mu'(x_1) \delta\mu'(x_2)}\,,\label{eq:d2Gamd2mu}\\[2ex]
    \Gamma^{(2)}_{k,\mu'\sigma}=&\frac{\delta^2\Gamma_{k}[\Phi, \mu']}{ \delta \mu'(x_1) \delta \sigma(x_2)}\,,\label{eq:d2Gamdmudsigma}
\end{align}
Note that in \eq{eq:d2Gamd2mu,eq:d2Gamdmudsigma} the field $\Phi$ is independent on $\mu'$, in contrast to the case in \Eq{eq:C2-d2Gamd2mu}. In \Eq{eq:C2-twoparts} $G_{\sigma}$ stands for the propagator of sigma meson, that reads
\begin{align}
    G_{\sigma}(p)=\frac{1}{p_0^2+\bm{p}^2+m_\sigma^2}\,,\label{}
\end{align}
with the sigma mass $m_\sigma$. The diagrammatic representation of \Eq{eq:C2-twoparts} is shown in \Fig{fig:ogogsg}. Note that the analog of the two-loop diagram in \Fig{fig:ogogsg} with the pion propagator in place of the sigma propagator is vanishing identically due to the Dirac structure. A similar equation as \Eq{eq:C2-twoparts} was also derived in the appendix in Ref.~\cite{Fu:2015naa}, see equation (D8) therein.

The two-point functions in \eq{eq:d2Gamd2mu,eq:d2Gamdmudsigma} are computed from their flow equations within the fRG approach, whose diagram representation is shown in \Fig{fig:op2}, where the RG time is defined as $t=\ln(k/\Lambda)$, with $\Lambda$ denoting a ultraviolet reference scale. The flow equations for $\Gamma^{(2)}_{k,\mu'\mu'}$ and $\Gamma^{(2)}_{k,\mu'\sigma}$ are presented in \Cref{app:flow-2point}.

\section{Numerical results}
\label{sec:num} 

%
\begin{figure}[t]
\includegraphics[width=0.45\textwidth]{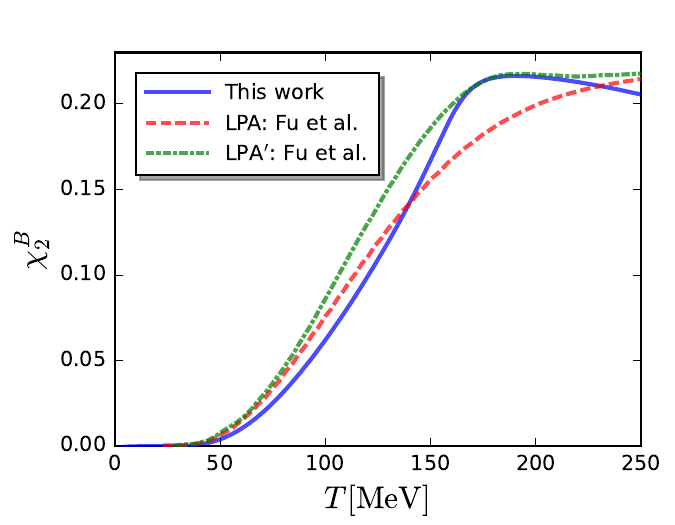}
\caption{Quadratic baryon number fluctuation as a function of temperature $T$ at vanishing baryon chemical potential. The result from this work is compared to those from Ref. \cite{Fu:2015naa} obtained within the $\mathrm{LPA}$ and $\mathrm{LPA}'$ approximations of the fRG approach.}
\label{fig:chi2}
\end{figure}
%

%
\begin{figure}[t]
\includegraphics[width=0.45\textwidth]{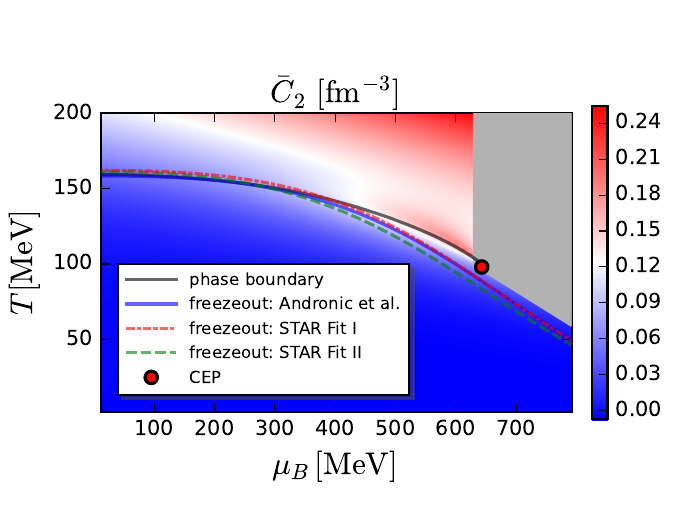}
\caption{Heatmap of $\bar C_2$ in \Eq{eq:C2bar} in the QCD phase diagram. The red dot denotes the critical end point. The black solid line represents the chiral crossover, and other lines represent three different freeze-out curves \cite{Fu:2021oaw}. The gray area depicts the region where the computation is inaccessible yet.}
\label{fig:C2bar}
\end{figure}
%

%
\begin{figure}[t]
\includegraphics[width=0.45\textwidth]{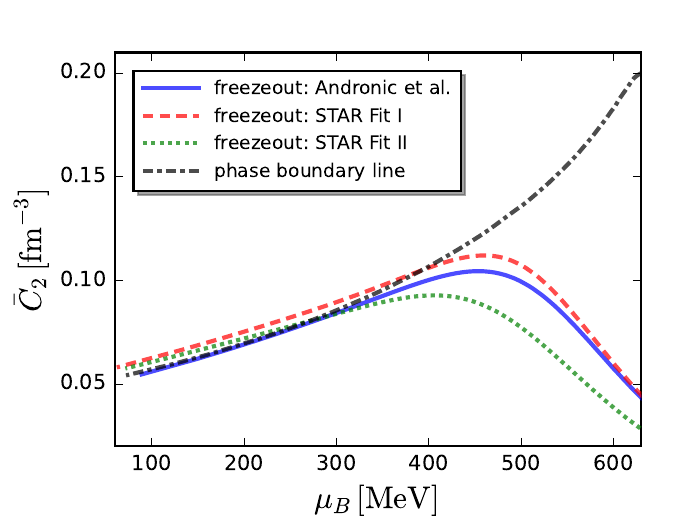}
\caption{$\bar C_2$ in \Eq{eq:C2bar} as a function of the baryon chemical potential on the phase boundary line and the three freeze-out curves.}
\label{fig:C2bar-muB}
\end{figure}
%

In order to calculate the flows of two-point functions $\Gamma^{(2)}_{k,\mu'\mu'}$ and $\Gamma^{(2)}_{k,\mu'\sigma}$ in \Fig{fig:op2}, we employ the Yukawa coupling and the masses of quarks and mesons calculated in \cite{Fu:2023lcm}, where the squared masses read
\begin{align}
    m^{2}_{q,k}&=\frac{1}{2}h^{2}_{k}\rho\,, \\[2ex]
    m^{2}_{\pi,k}&=V'_{k}(\rho)\,, \\[2ex]
    m^{2}_{\sigma,k}&=V'_{k}(\rho)+2\rho V''_{k}(\rho)\,.\label{}
\end{align}
for the quark, pion and sigma, respectively. Note that the renormalized quantities in \cite{Fu:2023lcm} are used as inputs to compute $C_2(p_0,\bm{p})$. Then by performing the sum over the Matsubara frequency $p_0$, one obtains the equal-time two-baryon density fluctuation correlation function $C_2(\tau=0,\bm{p})$.

The two-point baryon density correlation $C_2(\tau=0,\bm{p})$ with two typical values of momentum, $\bm{p}=0$ and $|\bm{p}|=\sqrt{2}/\sigma_d$, is depicted in the QCD phase diagram in \Fig{fig:C2}. Note that $|\bm{p}|=\sqrt{2}/\sigma_d$  corresponds to the momentum mode that gives the dominant contribution to the $\bar C_2$ integral in \Eq{eq:C2barII}. The red dot stands for the CEP, which is located at $(\mu_{B_{\mathrm{CEP}}},T_{_{\mathrm{CEP}}})=(643,98)$ MeV in the low energy effective field theory \cite{Fu:2023lcm}. The black solid line denotes the continuous chiral crossover, i.e., the phase boundary line. The other lines of different colors stand for three representative chemical freeze-out curves, which are introduced in \cite{Fu:2021oaw} and denoted by Andronic et al. \cite{Andronic:2017pug}, STAR Fit I, and STAR Fit II \cite{Adamczyk:2017iwn}, respectively, see \cite{Fu:2021oaw} for more details. 

One finds from \Fig{fig:C2} that there is a red band along the phase boundary radiated from the CEP, which indicates that the two-point baryon density correlation $C_2$ is enhanced by the critical fluctuations related to the CEP in the regime of red band. In order to see this more clearly, we show $C_2$ along the phase boundary and the freeze-out curves in \Fig{fig:C2-muB}, where the solid and dashed lines correspond to the spatial momentum $\bm{p}=0$ and $|\bm{p}|=\sqrt{2}/\sigma_d$, respectively. It is found that the fluctuation-driven enhancement for the $C_2$ is weakened with the increase of the momentum, since the dashed line results of $|\bm{p}|=\sqrt{2}/\sigma_d$ are systematically smaller than the solid line results of $\bm{p}=0$. Moreover, we can see that the density correlation along the phase boundary increases monotonically with the increase of $\mu_B$, while that on the chemical freeze-out curves increases firstly and then decreases. This arises from the fact that the freeze-out curves bend down and deviate from the phase boundary in the regime of large baryon chemical potential, which is also clearly observed in \Fig{fig:C2}, where the freeze-out curves deviate from the red band radiated from the CEP at large $\mu_B$. Therefore, it is expected that the effect of critical fluctuations from CEP on the two-point baryon density correlation on the freeze-out curves will not be significant.

The two-point baryon density correlation essentially can be related to the conventional baryon number fluctuations measured in experiments (proxied by the net-proton fluctuations) by
\begin{align}
    \chi_2^B=\frac{\langle(\delta N_B)^2\rangle}{V T^3}=\frac{C_2(\tau=0,\bm{p}=0)}{T^3}\,,\label{eq:chi2B}
\end{align}
with $\delta N_B=N_B-\langle N_B\rangle$ and $\langle \cdots \rangle$ denoting the ensemble average. Thus, $C_2(\tau=0,\bm{p}=0)$, i.e., the equal-time two-point baryon density correlation at vanishing momentum, is just the quadratic baryon number fluctuation $\chi_2^B$, if we neglect the factor $T^3$. 

In \Fig{fig:chi2}, the second-order baryon number fluctuation computed from \Eq{eq:chi2B} is compared with the results in \cite{Fu:2015naa}, where two different truncations are adopted: One is the local potential approximation (LPA) and the other is the local potential approximation with wave function renormalization denoted by $\mathrm{LPA}'$. In \cite{Fu:2015naa} the baryon number fluctuations are calculated by performing derivative of the pressure with respect to the baryon chemical potential. From \Fig{fig:chi2} one can see the quadratic baryon number fluctuation computed from the two-point baryon correlation is qualitatively consistent with the results in \cite{Fu:2015naa}, which indicates that the decomposition of the two-point baryon density correlation into two parts shown in \Fig{fig:ogogsg}, i.e., \Eq{eq:C2-twoparts}, is a reasonable approximation.

Note that the two-point baryon density correlation in \Eq{eq:C2-d2Gamd2mu} is the correlation function for the net-baryon number, while that in the coalescence model in \Eq{eq:C2bar} is the correlation function of positive baryon number. In order to bridge this difference, we adopt
\begin{align}
    C_2^+\approx \frac{\langle N_B^+\rangle}{\langle N_B^+\rangle+\langle N_B^-\rangle}C_2, \label{eq:C2plus}
\end{align}
where $N_B^+$ and $N_B^-$ stand for the positive baryon number and the anti-baryon number, respectively, which are both positive. $C_2^+$ is the two-point correlation for the positive baryon number. \Cref{eq:C2plus} is motivated from the Skellam distribution, where the second-order, or more generally even-order net baryon fluctuation can be expressed as a sum of particle and anti-particle fluctuations, which are proportional to their expected value of particle number \cite{Morita:2014fda, Sun:2018ozp}. Apparently, the factor on the right side of \Eq{eq:C2plus} approaches to unity in the regime of large baryon chemical potential, where one has $\langle N_B^+\rangle \gg \langle N_B^-\rangle$ and thus $C_2^+\sim C_2$. While in the other limit with $\mu_B=0$, one is left with $C_2^+\sim C_2/2$ due to $\langle N_B^+\rangle \sim \langle N_B^-\rangle$. The numbers of baryon and anti-baryon at a finite temperature and chemical potential are calculated as follows
\begin{align}
    \langle N_B^+\rangle=&\frac{2}{3}N_c N_f V\int \frac{d^3 {\bm{p}}}{(2 \pi)^3} \frac{1}{e^{(\sqrt{\bm{p}^2+m_q^2}-\mu) / T}+1}\,,\label{}\\[2ex]
    \langle N_B^-\rangle=&\frac{2}{3}N_c N_f V \int \frac{d^3 {\bm{p}}}{(2 \pi)^3} \frac{1}{e^{(\sqrt{\bm{p}^2+m_q^2}+\mu) / T}+1}\,,\label{}
\end{align}
with $N_f=2$ and $N_c=3$.

\Cref{fig:C2bar} shows our calculated $\bar C_2$ in \Eq{eq:C2bar} or \Eq{eq:C2barII} in the QCD phase diagram. Similar with the two-point baryon correlation function in \Fig{fig:C2}, there is an enhancement band radiated from the CEP along the phase boundary. In the same way, $\bar C_2$ is also depicted along the phase boundary and the chemical freeze-out curves in \Fig{fig:C2bar-muB}, where similar behaviors as the baryon density correlation in \Fig{fig:C2-muB} are observed: $\bar C_2$ increases monotonically with $\mu_B$ along the phase boundary, while it develops a peak structure along the freeze-out curves. In brief, the chemical freeze-out curves bend down and deviate from the enhancement region related to the CEP, which implies that the effect of critical fluctuations on the deuteron yield from density correlation is mild.

%
\begin{figure}[t]
\includegraphics[width=0.45\textwidth]{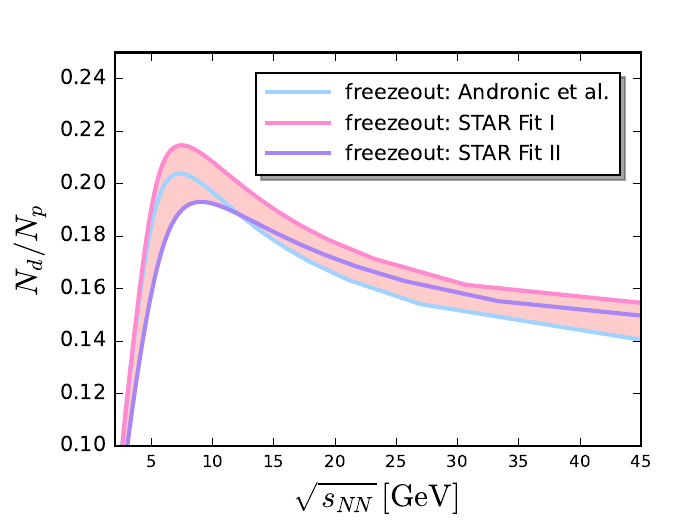}
\caption{Collision energy dependence of $N_d/N_p$ presented in \Eq{eq:Nd} along three distinct freeze-out curves.}
\label{fig:NdNp-SNN}
\end{figure}
%

%
\begin{figure}[t]
\includegraphics[width=0.45\textwidth]{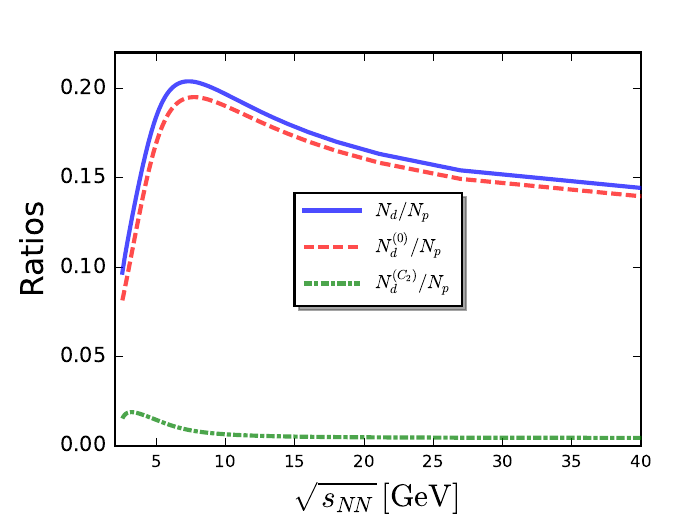}
\caption{Collision energy dependence of the $N_d/N_p$, $N_d^{(0)}/N_p$ and $N_d^{(C_2)}/N_p$ in \Eq{eq:Nd} along the chemical freeze-out curve denoted by Andronic et al. \cite{Andronic:2017pug}.}
\label{fig:NdNps-SNN}
\end{figure}
%

With the calculated $\bar C_2$ in \Fig{fig:C2bar}, one is able to calculate the deuteron yield arising from the density correlation through \Eq{eq:Ndc2}. Furthermore, we also compute the leading-order contribution to the deuteron yield without density correlations, i.e., $N_d^{(0)}$ in \Eq{eq:Nd0}, where $C_{np}$ is assumed to be vanishing. The proton and neutron densities at the chemical freeze-out are estimated as
\begin{align}
    \langle\rho_{p}\rangle\approx\langle\rho_{n}\rangle\approx\frac{1}{2}\frac{\langle N_B^+\rangle}{V},\label{eq:rhopn}
\end{align}
where the isospin symmetry is assumed. 

Note that in the experimental measurements during the evolution from the chemical freeze-out to the kinetic freeze-out, the deuteron yield $N_d$ is approximately conserved, while the proton yield $N_p$ increases significantly due to the strong decays of resonances. This enhancement can be described by 
\begin{align}
    N_{p,\mathrm{kin}} = R\,N_{p,\mathrm{che}},\label{eq:Np-kin}
\end{align}
where $N_{p,\mathrm{kin}}$ and $N_{p,\mathrm{che}}$ denote the proton number at the kinetic and chemical freeze-out, respectively. The enhancement factor $R\sim 3$ is found in the regime of low baryon chemical potential \cite{Vovchenko:2019pjl}. In this work $R=3$ is assumed throughout all the collision energies for simplicity. 

In \Fig{fig:NdNp-SNN} we show the ratio between the deuteron and proton yields as a function of the collision energy, calculated on the three different chemical freeze-out curves with the proton yield corrected with the relation in \Eq{eq:Np-kin}. The results of different freeze-out curves in turn can be used to estimate the errors stemming from the sensitivity on the freeze-out curve, and thus a band in between curves is depicted in \Fig{fig:NdNp-SNN}. One can see that as the collision energy is lowered down starting from the regime of high energy, the ratio increases firstly, and then begins to decrease at between $\sqrt{s_{NN}}=5$ and 10 GeV. In \Fig{fig:NdNps-SNN}, we show the contributions of different parts to the total deuteron yield, including the leading-order contribution $N_d^{(0)}$ in \Eq{eq:Nd0}, and that from the density correlation $N_d^{(C_2)}$ in \Eq{eq:Ndc2}. One can see that the leading-order contribution $N_d^{(0)}$ plays a dominant role, that is, $N_d^{(C_2)}$ is significantly smaller than $N_d^{(0)}$. This is reasonable, since in the regime of low collision energy, i.e., the region of large baryon chemical potential, the freeze-out curves deviate from the critical region, resulting in that the enhancement of the deuteron yield stemming from the critical fluctuations near the CEP is mild.

\section{Conclusions}
\label{sec:con}

Based on the nucleon coalescence model and a low energy effective field theory of quarks and mesons, we investigate the influence of QCD phase transition and critical fluctuations of the CEP on the deuteron yield within the fRG approach. The two-point baryon density correlation function is decomposed into a sum of two parts: The first part is the one-particle-irreducible (1PI) vertex of vector currents, and the other part is the two mixed 1PI vertices of vector and scalar currents connected by a sigma propagator. This decomposition allows us to compute the 1PI vertices by using their flow equations within the fRG approach.

It is found that the two-point baryon density correlation function as well as the resulting deuteron yield is enhanced in a narrow region radiated from the CEP along the phase boundary. We also compute the baryon density correlation and the resulting deuteron yield on the chemical freeze-out curves and the phase boundary line. It is found that both the baryon density correlation and the resulting deuteron yield on the phase boundary increase monotonically with the increasing baryon chemical potential, while those on the freeze-out curves develop a peak structure, which is attributed to the fact that the chemical freeze-out curves bend down and deviate from the enhancement region related to the CEP in the regime of large chemical potential.

The ratio between the deuteron and proton yields is calculated as a function of the collision energy. It is found that as the collision energy is lowered down starting from the regime of high energy, the ratio increases firstly and then decreases. The deuteron yield resulting from the two-point baryon correlation is small compared to the leading-order contribution, since the enhancement of the deuteron yield due to the critical fluctuations of CEP is mild as we have discussed above.

\section*{Acknowledgements}

We thank Kai-jia Sun, Yang-yang Tan, Shan-jin Wu and, in particular Shi Yin for discussions and comments. We also would like to thank the members of the fQCD collaboration \cite{fQCD} for collaborations on related projects. This work is supported by the National Natural Science Foundation of China under Contract Nos.\ 12447102, 12175030. Y.-r. Chen is supported by the Alexander von Humboldt foundation through Humboldt Research
Fellowship for postdocs.\\


\appendix

\section{Flow equation of two-point correlation functions}
\label{app:flow-2point}

The flow equations for $\Gamma^{(2)}_{k,\mu'\mu'}$ in \Eq{eq:d2Gamd2mu} and $\Gamma^{(2)}_{k,\mu'\sigma}$ in \Eq{eq:d2Gamdmudsigma} are given as follows:
\begin{align}
    \partial_{t}\Gamma^{(2)}_{k,\mu'\mu'}(p)&=-2N_cN_fkJ_{k,\mu'\mu'}(p)\,,\label{eq:d2Gamdmumuliu}\\[2ex]
    \partial_{t}\Gamma^{(2)}_{k,\mu'\sigma}(p)&=-2N_cN_fkJ_{k,\mu'\sigma}(p)\,,\label{eq:d2Gamdmusigmaliu}
\end{align}
where one has
\begin{align}
    J_{k,\mu'\mu'}(p)&=\int_{D_1}\frac{{d}^{3}\bm{q}}{\left(2\pi\right)^{3}}4k\Big[l_1+(1+F_1)l_2+(2E_{k,q}^2\nonumber\\[2ex]&+2m_{k,q}^2+p_0^2+2k^2F_1)l_3\Big]+\int_{D_2}\frac{{d}^{3}\bm{q}}{\left(2\pi\right)^{3}}4\Big[kl_1\nonumber\\[2ex]
    & +(k+F_2)l_4+\big((E_{k,q}^2+E_{p+q,q}^2+2m_{k,q}^2+p_0^2)k\nonumber\\[2ex]
    &+2k^2F_2\big)l_5\Big]\,,\label{eq:Jmu}\\[2ex]
    J_{k,\mu'\sigma}(p)&=-\frac{4km_{k,q} h_k}{\mathrm{i} p_0}\Big[\int_{D_1}\frac{{d}^{3}\bm{q}}{\left(2\pi\right)^{3}}   (l_1-l_2)\nonumber\\[2ex]
    &+ \int_{D_2}\frac{{d}^{3}\bm{q}}{\left(2\pi\right)^{3}}\big( l_1-l_4-(E_{k,q}^2 - E_{\bm{p}+\bm{q},q}^2)l_5\big)\Big]\,,\label{eq:Jmusigma}
\end{align}
with the integration domains  $D_1$  and $D_2$ defined by $\theta(k^{2}-(\bm{p}+\bm{q})^{2})\theta(k^{2}-\bm{q}^{2})$ and $\theta(-{k}^{2}+(\bm{p}+\bm{q})^{2})\theta(k^{2}-\bm{q}^{2})$, respectively. Here, we have
\begin{align} 
 {F}_{1}\equiv\frac{(\bm{p}+\bm{q})\cdot\bm{q}}{|\bm{p}+\bm{q}||\bm{q}|}\,,\qquad {F}_{2}\equiv\frac{(\bm{p}+\bm{q})\cdot\bm{q}}{|\bm{q}|}\,.\label{}
\end{align}
The threshold functions in \Eq{eq:Jmu} and \Eq{eq:Jmusigma} are given by
\begin{widetext}
\begin{align}
    l_{1} &\equiv T \sum_{n \in Z} \frac{1}{-\big((i q_0 - \mu)^2 - E_{k,q}^2\big)^2}\nonumber\\[2ex]
    & = -\frac{1}{4 E_{k,q}^3} \Big[ 1 - n_{F}(E_{k,q} + \mu) - n_{F}(E_{k,q} - \mu) + E_{k,q} \big( n_{F}'(E_{k,q} + \mu) + n_{F}'(E_{k,q} - \mu) \big) \Big] \,,\\[2ex]
   l_{2} &\equiv T \sum_{n \in Z} \frac{1}{\big( E_{k,q}^2 - (i q_0 + i p_0 - \mu)^2 \big) \big( (i q_0 - \mu)^2 - E_{k,q}^2 \big)} \nonumber\\[2ex]& = - \frac{1}{4 E_{k,q}^3 + E_{k,q} p_0^2} \Big( 1 - n_{F}(E_{k,q} + \mu) - n_{F}(E_{k,q} - \mu) \Big) \,, \\[2ex]
   l_{3} &\equiv T \sum_{n \in Z} \frac{1}{\big( E_{k,q}^2 - (i q_0 + i p_0 - \mu)^2 \big) \big( (i q_0 - \mu)^2 - E_{k,q}^2 \big)^2} \nonumber\\[2ex]
   &= \frac{(12 E_{k,q}^2 + p_0^2) \big( 1 - n_{F}(E_{k,q} + \mu) - n_{F}(E_{k,q}- \mu) \big)}{4 E_{k,q}^3 (4 E_{k,q}^2 + p_0^2)^2}  + \frac{1}{4 E_{k,q}^2} \bigg( \frac{n_{F}'(E_{k,q} + \mu)}{p_0^2 - 2\mathrm{i} E_{k,q}  p_0} + \frac{n_{F}'(E_{k,q} - \mu)}{p_0^2 + 2 \mathrm{i} E_{k,q} p_0} \bigg)\,,\\[2ex]
   l_{4} &\equiv T \sum_{n \in Z} \frac{1}{\big( E_{\bm{p}+\bm{q},q}^2 - (i q_0 + i p_0 - \mu)^2 \big) \big( (i q_0 - \mu)^2 - E_{k,q}^2 \big)}\nonumber \\[2ex]
   & = \frac{1 - 2 n_{F}(E_{k,q} - \mu)}{4 E_{k,q} \big( (E_{k,q} - \mathrm{i} p_0)^2 - E_{\bm{p}+\bm{q},q}^2 \big)} + \frac{1 - 2 n_{F}(E_{k,q} + \mu)}{4 E_{k,q} \big( (E_{k,q} + \mathrm{i} p_0)^2 - E_{\bm{p}+\bm{q},q}^2 \big)} + \frac{1 - 2 n_{F}(E_{\bm{p}+\bm{q},q} - \mu)}{4 E_{\bm{p}+\bm{q},q} \big( (E_{\bm{p}+\bm{q},q} + \mathrm{i} p_0)^2 - E_{k,q}^2 \big)} \nonumber\\[2ex]
   &+ \frac{1 - 2 n_{F}(E_{\bm{p}+\bm{q},q} + \mu)}{4 E_{\bm{p}+\bm{q},q} \big( (E_{\bm{p}+\bm{q},q} - \mathrm{i} p_0)^2 - E_{k,q}^2 \big)}\,,\\[2ex]
    l_{5} &\equiv T \sum_{n \in Z} \frac{1}{\big( E_{\bm{p}+\bm{q},q}^2 - (i q_0 + i p_0 - \mu)^2 \big) \big( (i q_0 - \mu)^2 - E_{k,q}^2 \big)^2}\nonumber\\[2ex]
    &= \frac{\big( E_{k,q}^2 + E_{\bm{p}+\bm{q},q}^2 - (2 E_{k,q} - \mathrm{i} p_0)^2 \big) \big( 1 - 2 n_{F}(E_{k,q} - \mu) \big)}{8 E_{k,q}^3 \big( (E_{k,q} - \mathrm{i} p_0)^2 - E_{\bm{p}+\bm{q},q}^2 \big)^2} + \frac{\big( E_{k,q}^2 + E_{\bm{p}+\bm{q},q}^2 - (2 E_{k,q} + \mathrm{i} p_0)^2 \big) \big( 1 - 2 n_{F}(E_{k,q} + \mu) \big)}{8 E_{k,q}^3 \big( (E_{k,q} + \mathrm{i} p_0)^2 - E_{\bm{p}+\bm{q},q}^2 \big)^2}\nonumber\\[2ex]
    & + \frac{1 - 2 n_{F}(E_{\bm{p}+\bm{q},q} - \mu)}{4 E_{\bm{p}+\bm{q},q} \big( (E_{\bm{p}+\bm{q},q} + \mathrm{i} p_0)^2 - E_{k,q}^2 \big)^2} + \frac{1 - 2 n_{F}(E_{\bm{p}+\bm{q},q} + \mu)}{4 E_{\bm{p}+\bm{q},q} \big( (E_{\bm{p}+\bm{q},q} - \mathrm{i} p_0)^2 - E_{k,q}^2 \big)^2} - \frac{ n_{F}'(E_{k,q} + \mu)}{4 E_{k,q}^2 \big( (E_{k,q} + \mathrm{i} p_0)^2 - E_{\bm{p}+\bm{q},q}^2 \big)}\nonumber \\[2ex]
    &- \frac{ n_{F}'(E_{k,q} - \mu)}{4 E_{k,q}^2 \big( (E_{k,q} - \mathrm{i} p_0)^2 - E_{\bm{p}+\bm{q},q}^2 \big)}\,,\label{}
\end{align}
\end{widetext}
with $p_0=2n\pi T~(n\in Z)$, $q_0=(2n+1)\pi T~(n\in Z)$. The effective quasi-particle energy $E_{k}$ and $E_{p+q}$ are defined as
\begin{align}
E_{k,q}&\equiv\sqrt{k^2+m^2_{q}}\nonumber\,,\\[2ex]
E_{p+q,q}&\equiv\sqrt{(\bm{p}+\bm{q})^2+m^2_{q}}\,.
\end{align}
The fermionic distribution function reads
\begin{align}
    n_F(x)=\frac{1}{e^{\frac{x}{T}}+1}\,.
\end{align}

In this work, we employ the $3d$ optimized, i.e., flat infrared regulators \cite{Litim:2000ci, Litim:2001up}, i.e.,
\begin{align}
    R_{k,\phi\phi}(q)&=\bm{q}^2 \,r_{B}(\bm{q}^2/k^2)\,, \\[2ex]
    R_{k,\bar q q}(q)&=\mathrm{i} \bm{\gamma} \cdot \bm{q} \,r_{F}(\bm{q}^2/k^2)\,,
\end{align}
with
\begin{align}
    r_B(x)&=(\frac{1}{x}-1)\theta(1-x)\,, \\[2ex]
    r_F(x)&=(\frac{1}{\sqrt{x}}-1)\theta(1-x)\,. 
\end{align}


\bibliography{ref-lib}%

\end{document}